\documentclass[11pt,twoside]{article}
\usepackage{fleqn,espcrc1}
\usepackage{amssymb}
\usepackage{amsmath}  
\newcommand{\rd}{{\rm d}}
\newcommand{\ri}{{\rm i}}
\newcommand{\re}{{\rm e}}
\newcommand{\ao}{{\hat A}}
\newcommand{\bo}{{\hat B}}
\newcommand{\Io}{{\hat I}}
\newcommand{\no}{{\hat N}}
\newcommand{\io}{{\hat J^\dagger}}
\newcommand{\jo}{{\hat J}}
\newcommand{\ko}{{\hat K}}
\newcommand{\co}{{\hat C}}
\newcommand{\so}{{\hat S}}
\newcommand{\cco}{{\hat C^2}}
\newcommand{\sso}{{\hat S^2}}
\newcommand{\ho}{{\hat K^\dagger}}
\newcommand{\hho}{{\hat K^{\dagger 2}}}
\newcommand{\kko}{{\hat K^{2}}}
\newcommand{\hmo}{\hat K^{\dagger \,m}}
\newcommand{\nno}{{\hat N^{2}}}
\newcommand{\xo}{{\hat X}}
\newcommand{\yo}{{\hat Y}}
\newcommand{\dd}{d}
\newcommand{\kk}{\kappa}

\newcommand{\ad}{\,{\rm ad}}
\newcommand{\sesum}{\!\subset \! \! \! \!\! \! \!+\,}
\begin{document}
\title{\bf \LARGE An algebraic solution of driven single band tight binding 
dynamics}
\author{\bf \large H.~J.~Korsch\thanks{{\it E-mail}: korsch@physik.uni-kl.de}
and S.~Mossmann
\\FB Physik, TU Kaiserslautern, D-67653 Kaiserslautern, Germany}

\maketitle

\begin{abstract}
\vspace*{4mm}
\noindent
{\bf Abstract:}\\
The dynamics of the driven single band tight binding model for Wannier-Stark
systems is formulated and solved 
using a dynamical algebra. This Lie algebraic approach is very convenient
for evaluating matrix elements and expectation values. 
A classicalization
of the tight binding model is discussed as well as some illustrating
examples of Bloch oscillations and dynamical localization effects.
It is also shown that a dynamical invariant can be constructed.\\
\end{abstract}


\ \\
\noindent
\section{Introduction}

The celebrated single band tight binding system
\begin{equation}
\label{Hamtb}
H=-\frac{\Delta}{4} \sum_{n=-\infty}^{+\infty} \big(\,
|n\rangle \langle n +1| + 
| n +1\rangle \langle n|\,\big)
+\dd F \sum_{n=-\infty}^{+\infty} n |n \rangle \langle n |
\end{equation}
models a space periodic system with period $\dd$ in a (possibly time dependent)
linear field. Here, $n$ numbers the sites and $|n\rangle$ are the Wannier states 
with \,$\langle n|n'\rangle=\delta_{nn'}$\,. In (\ref{Hamtb}) only nearest 
neighbor interactions are taken into account. In this model, the periodic field free system has only a single band with dispersion
relation
\begin{equation}
E(\kk)=-\frac{\Delta}{2}\cos \,(\kk \dd)
\label{Ekappa}
\end{equation}
where $\kk$ is the Bloch index and $\Delta$ is the band width. In the simplest case, the 
field $F$ is constant, a dc-field. More complicated is the combined 
ac-dc-system with time periodic driving; an often considered case is the
harmonic driving,\,$F(t)=F_0-F_1\cos (\omega t)$\,. 

The dynamics of the driven tight binding system is quite involved and,
despite of the large number of previous studies, of increasing
interest, in particular in view of the recent progress in studies
of the dynamics of ultracold atoms in standing wave
laser fields.
For more information, see \cite{Grif98,03bloch0} and the
references given there. It is well known that the tight binding system
allows an analytic treatment and various approaches have been proposed
(for early studies see \cite{Fuku73,Dunl86,Krie86}).
Quite generally, however, the derivations are quite tedious. Here, we
recommend a treatment based on the dynamical Lie algebra
\cite{79inv,81inv,88lie} which appears to be favorable
because of its generality and simplicity.
This approach allows a straightforward evaluation of the time evolution
operator, matrix elements, expectation values and dynamical invariants
by purely algebraic operations.

\section{The algebra}

The three operators $\no$, $\ko$, $\ho$ where $\no$ is hermitian and $\ko$ unitary
with commutation relations
\begin{equation}
\big[\ko,\no\big]=\ko \ ,\quad \big[\ho,\no\big]=-\ho \ ,\quad
\big[\ho,\ko\big]=0 
\label{comm}
\end{equation}
form a closed Lie algebra $\cal L$. This shift-operator algebra
\cite{sack58} is obviously different from the ubi\-quit\-ous
oscillator algebra $\big\{\hat n,\, \hat a,\, \hat a^\dagger\big\}$, but some
features are similar (see also \cite{Ride92} for a discussion of the
more general quantum boson algebra which contains both algebras as limiting
cases). 
 
The dynamics generated 
by the hermitian
Hamiltonian
\begin{equation}
\hat H=G(t)\,\big(\ko+\ho\big)+F(t)\no
\label{ham}
\end{equation}
with real valued, possibly time dependent functions $F$ and $G$ can be conveniently
studied by algebraic techniques.

A realization is the tight binding model (\ref{Hamtb}) with 
\begin{equation}
\no = \sum_{n=-\infty}^{+\infty} n |n \rangle \langle n |
\quad , \quad
\ko = \sum_{n=-\infty}^{+\infty} |n\rangle \langle n +1| \,.
\quad , \quad
\ho = \sum_{n=-\infty}^{+\infty} |n+1\rangle \langle n| \,.
\end{equation}
It should be noted, however, that this algebra appears also in different 
context \cite{sack58}
and therefore some general considerations seem to be appropriate. First, one can easily
show that $\ko$ and $\ho$ act on the eigenstates of $\no$, 
$\no |n\rangle =n |n\rangle $, as shift- or ladder-operators: 
\begin{equation}
\ko |n\rangle =|n-1\rangle \ , \quad \ho |n\rangle =|n+1\rangle\,, 
\label{ladder}
\end{equation}
which fixes the
eigenvalues of $\no$ at $n_0+n$ with $n\in \mathbb{Z}$ up to an arbitrary value of $n_0$
in the unit interval. This value can be fixed if one considers the algebra as
a subalgebra of a bigger one by adding an antiunitary operator representing time
inversion. This leads to the two possible cases of $n_0=0$ (bosonic) or  $n_0=1/2$ 
(fermionic). See \cite{Kowa96} and references given there for more details. Here
we are interested in the bosonic case, i.e.
\begin{equation}
\no\,|n\rangle = n |n\rangle \ , \quad n\in \mathbb{Z}\,.
\label{neigen}
\end{equation}
In context of the tight binding system, $\no$ is a
'position operator': the expectation value
$\langle N\rangle=\langle\psi|\no|\psi\rangle$ is the mean position
on the lattice and $p_n=|\langle n|\psi\rangle|^2$ is
the population probability of the 'lattice site' at position $n$.
This is the physical system we have in mind. It should be noted, however,
that the same algebra appears in different contexts as, e.g., for the plane
rotor $[\jo_z,\jo_\pm]=\pm \jo_\pm$, $[\jo_+,\jo_-]=0$\,.

The eigenvectors of $\ko$ with eigenvalues $\re^{\ri \kk}$ are
\,$|\kk \rangle = \frac{1}{ \sqrt{2\pi}} \sum_{n=-\infty}^{+\infty}\re^{\ri n\kk} |n\rangle$\,:
\begin{equation}
\ko\,|\kk \rangle = {\textstyle \frac{1}{\sqrt{2\pi}}} \sum_{n=-\infty}^{+\infty}\re^{\ri n\kk} |n-1\rangle 
={\textstyle \frac{1}{\sqrt{2\pi}}} \,\,\re^{\ri \kk}\!\!\sum_{m=-\infty}^{+\infty}\re^{\ri m\kk} |m \rangle 
= \re^{\ri \kk}\,|\kk\rangle\,.
\label{eigvkk}
\end{equation}
These 'Bloch states' are $2\pi$ periodic and normalized as 
\begin{equation}
\langle \kk |\kk'\rangle = \sum_{n=-\infty}^{+\infty}\delta(\kk -\kk'-2\pi n)
=\delta_{2\pi}(\kk-\kk')
\label{kknorm}
\end{equation}
where $\delta_{2\pi}$ is the $2\pi$-periodic comb function.

The representation of the operator $\no$ in this basis is
\begin{equation}
\langle \kk |\no|\kk'\rangle=\delta_{2\pi}(\kk-\kk')\,\ri\,\frac{\rd \ }{\rd \kk}\,.
\label{no-kk}
\end{equation}

The algebra ${\cal L}=\{\ko,\ho,\no\}$ has the radical 
${\cal R}=\{\ko,\ho\}$\,,
the simple part ${\cal S}=\{\no\}$, and can be decomposed into the semidirect sum 
\,${\cal L}={\cal R} \sesum{\cal S}$\, as, e.g., described in \cite{88lie}.
For a Hamiltonian \,$H=H_S+H_R$ with
$H_R\in {\cal R}$ and $H_S\in{\cal S}$,
the time evolution operator can be factorized:
\begin{equation}
\hat U=\hat U_S\,\hat U_R
\label{USR1}
\end{equation}
with
\begin{equation}
\ri \hbar \,\frac{\rd \hat U_S}{\rd t}=\hat H_S\,\hat U_S \ ,  \quad
\ri \hbar\,\frac{\rd \hat U_R}{\rd t}=\big(\hat U_S^{-1}\hat H_R\hat
U_S\big)\ \hat U_R\,.
\label{USR2}
\end{equation}
(Moreover, $U_S$ can be factorized into simple parts if ${\cal S}$ is only 
semisimple; for more details see \cite{88lie}.)

This product decomposition has several advantages in comparison with the
pure exponential solution, in particular it provides a global solution 
\cite{Wei63-1,Wei63-2} and the calculation
of expectation values and  matrix elements is simplified
as will become clear later on.

The initial step of any application is the evaluation of all necessary
$\hat \Gamma$-evolved operators $\hat A$ of interest, i.e.
\begin{eqnarray}
&&\re^{z \ad \hat \Gamma}\hat A:=\re^{z\hat \Gamma}\hat A\,\re^{-z\hat \Gamma}
\nonumber \\
&&=\hat A+z\,\big[\,\hat \Gamma,\hat A \,\big]
+\frac{z^2}{2!}\,\big[\,\hat \Gamma,\big[\,\hat \Gamma,\hat A \,\big] \,\big]
+\frac{z^3}{3!}\,\big[\,\hat \Gamma,\,\big[\,\hat \Gamma,\big[\,\hat \Gamma,\hat A \,\big] \,\big] \,\big]
+ \,\ldots
\label{zevol}
\end{eqnarray}
with $z\in \mathbb{C}$ for all $\Gamma \in {\cal L}$. Trivially
we have \,$\re^{z \ad \hat \Gamma}\hat \Gamma =\hat\Gamma$\,.
Here we certainly need
the evolved operators of our algebra ${\cal L}=\{\ko,\ho,\no\}$\,.
Because $\ko$ and $\ho$ commute, we have
\begin{equation} 
\re^{z \ad \ko} F(\ho)=F(\ho) \ , \quad
\re^{z \ad \ho}F(\ko)=F(\ko)
\label{zfk}
\end{equation}
and the nontrivial expressions are
\begin{equation} 
\re^{z \ad \ko}\no=\no +z\ko\ , \ 
\re^{z \ad \no}\ko=\re^{-z}\,\ko \ ,\ 
\re^{z \ad \ho}\no=\no-z\ho \ ,\ 
\re^{z \ad \no}\ho= \re^{+z}\,\ho
\label{zn-kh}
\end{equation}
which can be easily obtained from (\ref{zevol}) using (\ref{comm}).

\section{Time evolution operator}

For the tight binding Hamiltonian (\ref{ham}), where for simplicity we
introduce the notation $g_t=G(t)/\hbar$, $f_t=F(t)/\hbar$, the simple part of 
the time evolution is
\,$\ri \,\dot{\hat U}_S=f_t\no\,\hat U_S $\,
with solution
\begin{equation}
\hat  U_S(t)=\re^{-\ri \eta_t\no} \ , \quad \eta_t=\int_0^t\! f_\tau \,\rd \tau 
\label{USt}
\end{equation}
and the remaining  equation of motion 
\,$\ri \,\dot{\hat U}_R=\big(\hat U_S^{-1}\hat H_R\,\hat U_S\big)\,\hat U_R $\, for
the radical part can be solved in a second step. 
Using the relation (\ref{zn-kh}) we find
\begin{equation}
\hat U_S^{-1}\hat H_R\,\hat U_S=\re^{\ri \eta_t \ad \no}\hat H_R
=\hbar g_t\,\big(\re^{-\ri \,\eta_t}\ko
+\re^{+\ri \,\eta_t} \ho\big)
\end{equation}
and therefore
\begin{equation}
\hat U_R(t)=\re^{-\ri\,(\chi_t\ko+\chi_t^*\ho)}\ , 
\quad \chi_t=\int_0^t\! g_\tau \,\re^{-\ri \,\eta_\tau}\,\rd \tau 
\label{URt}
\end{equation}
and finally
\begin{equation}
\hat U(t)=\hat U_S(t)\,\hat U_R(t)=\re^{-\ri \eta_t\no}\,
\re^{-\ri\,\chi_t\ko} \,\re^{-\ri\,\chi_t^*\ho }\,, 
\label{Ut}
\end{equation}
the Wei-Norman product form of the time evolution operator
\cite{88lie,Wei63-1,Wei63-2} which is, in fact, a version of the
so-called momentum gauge in this case.

Furthermore, a series expansion in powers of the ladder operator
will be useful, which can be obtained  by means of the generating function
for the Bessel functions
\begin{equation}
{\textstyle\re^{u\,( \hat B -\hat B^{-1})}}=\sum_{n=-\infty}^{+\infty}J_n(2u)\,\hat B^n\,.
\label{genfun}
\end{equation} 
Identifying \,$\hat B=\re^{-\ri(\phi_t+\pi/2)}\ko$\, where 
\,$\chi_t=|\chi_t|\,\re^{-\ri \phi_t}$\,, equation (\ref{URt}) can
be rewritten as
\begin{equation}
\hat U_R(t)=\sum_{n=-\infty}^{+\infty}J_n(2|\chi_t|)\,
\re^{-\ri n(\phi_t+\pi/2)}\,\ko^n\,.
\label{URt-exp}
\end{equation}

In various applications, matrix elements of the time evolution operator 
are required. Making use of
\begin{equation}
\langle \kk|\re^{-\ri u\no}| \kk'\rangle
=\sum_n\langle \kk| n\rangle\re^{-\ri u n}\langle n| \kk'\rangle
=\frac{1}{2\pi}\sum_n\re^{\ri n(\kk' -\kk-u)}=\delta_{2\pi}(\kk' -\kk -u)\,,
\label{eNkappa}
\end{equation}
the matrix elements in the Bloch wave basis can be directly read off from
(\ref{Ut}): 
\begin{equation}
\langle \kk|\hat U(t)|\kk'\rangle=
\delta_{2\pi}(\kk' -\kk-\eta_t)\,\re^{-2\ri |\chi_t|\cos(\kk'-\phi_t)}
\,. 
\label{Utkappa}
\end{equation}
Matrix elements of the propagator (\ref{Ut}) in the basis $|n\rangle$ 
follow immediately from (\ref{URt-exp}) and the ladder
property $\ko^n|n'\rangle=|n'-n\rangle$:
\begin{equation}
U_{nn'}(t)=\re^{-\ri (n'-n)(\phi_t+\pi/2) -\ri n\eta_t}\,J_{n'-n}(2\,|\chi_t|) 
\label{Utn}
\end{equation}
which coincides, of course, with the result
derived many years ago by Dunlap and Kenkre \cite{Dunl86}).

For completeness, we should also state the explicit results for the most
frequently studied cases:\\[2mm]
(1) For time independent
functions $g_t=g_0$ and $f_t=f_0$ the integrals in (\ref{USt}) and (\ref{URt}) 
yield
\begin{equation}
\eta_t=f_0t \ , \quad
\chi_t=\frac{2g_0}{f_0}\,\re^{-\ri f_0t/2}\,\sin(f_0t/2)\,.
\label{eta-chi-0}
\end{equation}
Note that at time $t=T_B=2\pi/f_0$ the evolution operator is equal to
the identity
\begin{equation}
\hat U(T_B)=\re^{-\ri 2\pi \no}=\hat I 
\end{equation}
because of $\re^{-\ri 2\pi \no}\, |n\rangle = \re^{-\ri 2\pi n}\,|n\rangle=
|n\rangle$\,. Therefore the dynamics is periodic with the Bloch period $T_B$
and Bloch frequency $\omega_B=f_0$.
It is also of interest to compare the product form (\ref{Ut}) of the
propagator with the pure exponential one which is trivial in
this case, namely
\begin{equation}
\hat U(t)=\re^{-\ri \hat H t/\hbar}=
\re^{-\ri \,(g_0\ko+g_0\ho+f_0\no)t}\,.
\label{Utexp}
\end{equation}
The non-obvious identity between (\ref{Utexp}) and (\ref{Ut})
becomes clear in view of the generalized Baker-Campbell-Hausdorff
formula \cite{sack58}
\begin{equation}
\re^{\alpha(\xo+\beta \yo)}=\re^{\beta(\re^\alpha -1) \yo}\,\re^{\alpha\xo}
=\re^{\alpha\xo}\,\re^{\beta (1-\re^{-\alpha})\yo}
\end{equation}
 for shift-operators $\xo$, $\yo$ with commutator
\,$[\xo,\yo]=\yo$\,.\\[2mm]
(2) For harmonic driving,
\begin{equation}
f_t=f_0-f_1\cos (\omega t)\ , \quad g_t=g_0 \,,
\label{fof1}
\end{equation}
we have
\begin{equation}
\eta_t=f_0t-\frac{f_1}{\omega}\,\sin (\omega t)
\end{equation}
and -- using again the Bessel expansion (\ref{genfun}) --
\begin{eqnarray}
\chi_t&=&g_0\int_0^t\!\rd \tau\,\re^{-\ri f_0\tau+\ri\frac{f_1}{\omega}\,\sin (\omega \tau)}
=g_0\sum_{\nu=-\infty}^\infty J_\nu \big({\textstyle \frac{f_1}{\omega}}\big)\int_0^t\!\rd \tau\,\re^{-\ri \omega_\nu \tau}
\nonumber \\
&=&2g_0 \sum_{\nu=-\infty}^\infty J_\nu \big({\textstyle \frac{f_1}{\omega}}\big)
\,\frac{1}{\omega_\nu}\,\re^{-\ri\,\omega_\nu t/2}\sin (\omega_\nu t/2)
\ , \quad \omega_\nu=\omega_B-\nu \omega \ne 0\,.
\label{besselint}
\end{eqnarray}
This is an oscillating function of time. For resonant 
driving,
\begin{equation}
\omega_B=n\,\omega \ ,\quad n=1,\,2,\,\ldots\,,
\end{equation}
the integration of the $n$th term in the sum (\ref{besselint}) yields a linearly growing
term, which dominates the oscillating rest of the sum for long times, i.e.~we have
\begin{equation}
\chi_t\approx \gamma_nt/2 \ ,\quad 
\gamma_n=2\,g_0 \,J_n \big({\textstyle \frac{f_1}{\omega}}\big)\,.
\label{Gamma}
\end{equation}
Later on, in section \ref{sec-dynamics}, some consequences of this resonant behaviour
will be discussed.\\[2mm]
(3) The general case of a combined dc- ac-system can be treated in a similar
manner. Let us consider the case
\begin{equation} 
f_t=f_0+\tilde f_t\ , \quad \tilde f_{t+T}=\tilde f_t\ , \quad g_{t+T}=g_t 
\label{ac-cd}
\end{equation}
with $f_0$ chosen according to $\int_0^T\tilde f_t \,\rd t=0$\,. Again we consider the case of
resonant driving, $T=nT_B$, with $T_B=2\pi/\omega_B$\,, $\omega_B=f_0$\,. 
Fourier expansion of the periodic part of the force
\begin{equation} 
\tilde f_t = \sum_{\mu = -\infty}^{+\infty} b_\mu \,\re^{\ri \mu \omega t}\ ,\quad b_0=0,
\label{fourier1}
\end{equation}
with $\omega=2\pi/T$ yields
\begin{equation} 
\eta_t = f_0t+
\sum_{\mu \ne 0} \frac{b_\mu}{\ri \mu \omega}
\,\big[\,\re^{\ri \mu \omega t} -1\big] =\omega_Bt+\tilde \eta_t
\label{etatilde}
\end{equation}
where  $\tilde \eta_t$ is $T$-periodic. A second Fourier expansion
\begin{equation} 
g_t\,\re^{-\ri \tilde \eta_t}= 
\sum_{\nu = -\infty}^{+\infty} a_\nu \,\re^{\ri \nu \omega t}
\ ,\quad a_\nu={\textstyle \frac{1}{2\pi}}\int_0^Tg_t\,\re^{-\ri \nu \omega t
-\ri \tilde \eta_t}\,\rd t\,,
\label{fourier2}
\end{equation}
allows the evaluation of $\chi_t$\,:
\begin{equation} 
\chi_t=\int_0^t g_\tau\,\re^{-\ri \eta_\tau}\,\rd \tau
=\int_0^t g_\tau\,\re^{-\ri \omega_B\tau -\ri \tilde \eta_\tau}\,\rd \tau
=a_nt+\sum_{\nu \ne n} \frac{a_\nu}{\ri \omega_\nu}
\,\big[\,1-\re^{-\ri \omega_\nu t}\big]
\end{equation}
with
\begin{equation} 
\omega_\nu=\omega_B-\nu \omega\,.
\end{equation}
Note that this is again a sum of a linear growing and a $T$-periodic part:
\begin{equation} 
\chi_t=a_nt+\tilde \chi_t \ ,\quad \tilde \chi_{t+T}=\tilde \chi_t\,.
\label{chitilde}
\end{equation}
The coefficients of two Fourier expansions (\ref{fourier1}) and
(\ref{fourier2}) are, of course, not unrelated for constant $g_0$. Let us
confine ourselves here for simplicity to the case of a symmetric resonant
driving:
\begin{equation} 
f_t=f_0+\sum_{m=1}^\infty f_m\cos (m\omega t) \ ,\quad \omega_B=n\omega\,,
\end{equation}
with -- inserting $f_0=\omega_B$ --
\begin{equation} 
\eta_t=\omega_Bt+\sum_{m=1}^\infty\beta_m\sin (m\omega t) \ ,\quad
\beta_m=f_m/m\omega\,.
\end{equation}
Using now the generating function for the infinite-variable
Bessel functions \cite{Lore95} (for a recent application
of these little known functions to a two-dimensional tight binding
system see \cite{02tb2d}),
\begin{equation} 
\exp \Big(\ri \sum_{m=1}^\infty \beta_m\sin mu \Big) = \sum_{\nu =-\infty}^{+\infty}
J_\nu(\{\beta_m \})\,\re^{\ri \nu u}\,,
\end{equation}
the remaining integral for $\chi_t$ can be evaluated analytically:
\begin{equation}
\chi_t=g_0\int_0^t \re^{-\ri \eta_\tau}\,\rd \tau 
=g_0J_n (\{\beta_m \})t + 2g_0 \sum_{\nu\ne n} J_\nu (\{\beta_m \})
\,\frac{1}{\omega_\nu}\,\re^{-\ri\,\omega_\nu t/2}\sin (\omega_\nu t/2)\,.
\label{inf-besselint}
\end{equation}
Similar to single frequency driving, dynamical localization effects
(see section \ref{sec-dynamics}) 
may be observed for system parameters leading to a zero of the 
infinite-order Bessel function $J_n (\{\beta_m \})$. This deserves future
studies.

\section{Quasienergies}

In the case of a combined dc- and time periodic ac-driving (\ref{ac-cd})
under resonance conditions $T=nT_B$, the time evolution operator (\ref{Ut})
over a period $T$ simplifies. Inserting (\ref{etatilde}) and (\ref{chitilde}) 
with  $\tilde \eta_T=\tilde \eta_0=0$ and $\tilde \chi_T=\tilde \chi_0=0$, 
we find
\begin{equation}
\hat U(T)=\re^{-\ri \eta_T\no}\,
\re^{-\ri\,\chi_T\ko} \,\re^{-\ri\,\chi_T^*\ho }
=\re^{-\ri\,a_nT\,\ko} \,\re^{-\ri\,a_n^*T\,\ho } 
\label{UT}
\end{equation}
which commutes with $\ko$ and allows the construction of 
simultaneous eigenstates
\begin{equation}
\ko \,|\psi_\kk(T)\rangle = \re^{\ri \kk }\,|\psi_\kk(T)\rangle \,.
\end{equation}
\begin{equation}
\hat U(T) \,|\psi_\kk(T)\rangle =\re^{-\ri\,a_nT\, \re^{+\ri \kk} 
-\ri\,a_n^*T\,\re^{-\ri \kk} }\,  |\psi_\kk(T)\rangle 
=\re^{-\ri \varepsilon_\kk T}\,|\psi_\kk(T)\rangle \,.
\end{equation}
The quasienergies $\varepsilon_\kk$ are identified as
\begin{equation}
\varepsilon_\kk =a_n \re^{+\ri \kk}+a_n^* \re^{-\ri \kk}=2|a_n|\cos (\kk+\varphi)\,,
\label{epsilon}
\end{equation}
the dispersion relation for the quasienergy. Here $4|a_n|$ is the width of the
quasienergy band and 
$\varphi$, the phase of the Fourier coefficient $a_n=|a_n|\,\re^{\ri \varphi}$,
is zero if $g_t$ and  $\tilde f_t$ are symmetric in time. 

It is also of interest to construct explicitly the time dependent
quasienergy (or Floquet) states. As can be easily seen, the states
\begin{equation} 
|\psi_\kk(t)\rangle = \hat U(t)\,|\kk\rangle=
{\textstyle \frac{1}{\sqrt{2\pi}}}
\sum_n \re^{\ri n \kk_t -\ri (\chi_t\re^{+\ri \kk} 
+\chi_t^*\re^{-\ri \kk})}|n\rangle
\ , \quad \kk_t=\kk-\eta_t
\label{quasistates}
\end{equation}
are solutions of the time dependent Schr\"odinger equation
\begin{equation} 
\big(\ri \,\frac{\partial \ \,}{\partial t} - \frac{1}{\hbar}\,\hat H\big)\,
|\psi_\kk(t)\rangle =0
\end{equation}
and, simultaneously, eigenstates of $\ko$\,:
\begin{equation}
\ko \,|\psi_\kk(t)\rangle = \re^{\ri \kk_t }\,|\psi_\kk(t)\rangle \,,
\end{equation}
the so-called Houston states.
Using again (\ref{etatilde}) and  (\ref{chitilde})), we have
\begin{eqnarray} 
|\psi_\kk(t+T)\rangle &=&
{\textstyle \frac{1}{\sqrt{2\pi}}}
\sum_n \re^{\ri n (\kk- \eta_{t+T}) -\ri (\chi_{t+T}\re^{+\ri\kk}
+\chi_{t+T}^*\re^{-\ri \kk})}|n\rangle
\nonumber \\[1mm]
&=&\re^{-\ri \omega_BT-\ri(a_n\re^{+\ri\kk}+a_n^*\re^{-\ri\kk})T}\,|\psi_\kk(t)\rangle
=\re^{-\ri\varepsilon_\kk T}\,|\psi_\kk(t)\rangle\,.
\end{eqnarray}
Therefore the state
\begin{equation} 
|u_\kk(t)\rangle = \re^{+\ri \varepsilon_\kk t} \,|\psi_\kk(t)\rangle
\end{equation}
with $\varepsilon_\kk$ given in (\ref{epsilon})
is $T$-periodic, \,$|u_\kk(t+T)\rangle =|u_\kk(t)\rangle $\, and
solves 
\begin{equation} 
\big(\ri \,\frac{\partial \ \,}{\partial t} - \frac{1}{\hbar}\,\hat H\big)\,
|u_\kk(t)\rangle =\varepsilon |u_\kk(t)\rangle \,,
\end{equation}
i.e.~it is a quasienergy state and $\varepsilon$ is the quasienergy.
From (\ref{quasistates}) we see that the Floquet states extend
over the whole lattice.
As a final remark, we also note the obvious identity 
\,$|u_\kk(T)\rangle =|\kk \rangle$\,.

\section{Expectation values}

The time dependence of expectation values follows immediately \cite{88lie}
from the relations (\ref{zfk})--(\ref{zn-kh}):
\begin{eqnarray} 
\ko (t)&=&\hat U^{-1}(t)\,\ko \,\hat U(t)
= \re^{\ri \chi_t^*\ho}\,\re^{\ri \chi_t\ko}\,\re^{\ri \eta_t\no}\,\ko\,
\re^{-\ri \eta_t\no}\,\re^{-\ri \chi_t\ko}\,\re^{-\ri \chi_t^*\ho}
\nonumber \\[1mm]
&=& 
\re^{\ri \chi_t^*\ho}\,\re^{\ri \chi_t\ko}\,\re^{-\ri \eta_t}\,\ko\,
\re^{-\ri \chi_t\ko}\,\re^{-\ri \chi_t^*\ho}
= \re^{-\ri \eta_t}\,\ko 
\label{Kt}
\end{eqnarray}
and therefore 
\begin{equation} 
\langle \ko \rangle_t =\re^{-\ri \eta_t}\,\langle \ko \rangle_0
=\re^{\ri (\kk-\eta_t)}\,|K| 
\label{Ktav}
\end{equation}
with $\langle \ko \rangle_0=K=|K|\,\re^{\ri \kk}$\,. From
\begin{equation} 
\ko^2 (t)= \re^{-2\ri \eta_t}\,\ko^2 
\quad \textrm{and} \quad
 \langle \ko^2 \rangle_t =\re^{-2\ri \eta_t}\,\langle \ko^2 \rangle_0\,,
\label{Kt2av}
\end{equation}
we see that, up to a phase factor, the variance is constant:
\begin{equation}
\Delta^2_K(t)=\big|\langle \kko \rangle_t-\langle \ko \rangle_t^2\big|=
\Delta^2_K(0)\,.
\label{varianceK}
\end{equation}
The time dependence of the position operator is a bit more interesting:
\begin{eqnarray} 
\no (t)&=& 
\re^{\ri \chi_t^*\ho}\,\re^{\ri \chi_t\ko}\,\no\,
\re^{-\ri \chi_t\ko}\,\re^{-\ri \chi_t^*\ho}
=\re^{\ri \chi_t^*\ho}\,\big( \no+\ri \chi_t\ko\big)\,
\re^{-\ri \chi_t^*\ho}\nonumber \\[1mm]
&=& \no +\ri \big(\chi_t\ko-\chi_t^*\ho\big)
\label{Nt}
\end{eqnarray}
and therefore, using $\chi_t=|\chi_t|\re^{-\ri \phi_t}$,
\begin{eqnarray} 
\langle \no \rangle_t 
&=&\langle \no \rangle_0 +\ri\,\big(\,\chi_t\,\langle \ko\rangle_0 - 
\chi_t^*\,\langle \ho\rangle_0\,\big) 
=\langle \no \rangle_0+2|K|\,|\chi_t|\,\sin(\phi_t-\kk). 
\label{Ntav}
\end{eqnarray}
Introducing
the anti-commutator \,$\jo=\no\ko+\ko\no=[\no,\ko]_+$\,, the time evolution of $\nno$ is 
\begin{eqnarray} 
&&\no^2 (t)
=\big(\,\no +\ri \,(\chi_t\ko-\chi_t^*\ho)\,\big)^2
=\nno\!+\!\ri\,(\,\chi_t\jo\!-\!\chi_t^*\io\,)\!-\!\chi_t^2\kko\!-\!\chi_t^{*\,2}\hho
+\!2\,|\chi_t|^2 
\label{N2t}
\end{eqnarray}
and, with 
\begin{eqnarray}
\langle \jo\,\rangle_0=J=|J|\,\re^{\ri \mu}\quad ,\quad
\langle \kko\rangle_0=L=|L|\,\re^{\ri \nu}\,,
\end{eqnarray}
the expectation value
evolves as
\begin{equation} 
\langle \nno \rangle_t 
=\langle \nno \rangle_0+2\,|J||\chi_t|\sin(\phi_t-\mu)
+2\,|\chi_t|^2\big(1-|L|\cos (2\phi_t-\nu)\,\big)\,. 
\label{Ntav2}
\end{equation}
Finally, the time evolution of the variance is given by
\begin{eqnarray}
\Delta^2_N(t)&=&\langle \nno \rangle_t\! -\!\langle \no \rangle_t^2 \nonumber\\[1mm]
&=&\Delta^2_N(0)+2|\chi_t|^2\big\{1\!-\!|L|\cos
(2\phi_t\!-\!\nu)-2|K|^2\sin^2(\phi_t\!-\!\kk)\big\}
\nonumber\\[1mm]
&&\quad + 2\,|\chi_t| \big\{\, 2\langle\no\rangle_0|K|\sin(\phi_t-\kk)+|J| \sin(\phi_t -\mu)\,\big\}\,.
\label{varianceN}
\end{eqnarray}
The dynamics of the expectation values therefore depends on three complex coherence
parameters which are explicitly
\begin{equation}
K=\sum_n c_{n-1}^*c_n \ , \quad
J=\sum_n (2n-1)\,c_{n-1}^*c_n \ , \quad
L=\sum_n c_{n-2}^*c_n
\label{KJL}
\end{equation}
if the initial normalized state is specified as 
\,$|\psi\rangle=\sum_nc_n\,|n\rangle$.

Sometimes it may be more convenient to replace the
unitary shift operators $\ko$ and $\ho$ by the hermitian operators
$\co$ and $\so$,
\begin{equation}
\ko =\co+\ri\,\so \ , \quad \ho =\co-\ri\,\so \,,
\label{CS}
\end{equation} 
whose expectation values allow a direct interpretation.
Clearly, also these
operators commute and the commutators with the position operator are
\begin{equation}
\big[\co,\no\,\big]=\ri \,\so \ ,\quad \big[\so,\no\,\big]=-\ri\, \co \,.
\label{commCS}
\end{equation}
Rewriting $\chi_t$ defined in equation (\ref{URt}) as
$2\chi_t=u_t-\ri v_t$ with
\begin{equation}
u_t=2\int_0^t\! g_\tau \,\cos \eta_\tau\,\rd \tau 
\ ,\quad
v_t=2\int_0^t\! g_\tau \,\sin \eta_\tau\,\rd \tau 
\label{utvt}
\end{equation}
the solution for the position operator in (\ref{Nt}) and (\ref{Ntav})
is
\begin{equation} 
\no (t)=\no + v_t\co- u_t\so
\label{NtCS}
\end{equation}
\begin{equation}
\langle \no \rangle_t =\langle \no \rangle_0
+v_t\langle\co\rangle_0- u_t\langle\so\rangle_0
\label{NtavCS}
\end{equation}
and, with \,$\jo=[\no,\co]_++\ri\,[\no,\so]_+$\,,
the time evolution of $\nno$ in (\ref{N2t}) and (\ref{Ntav2}) 
is rewritten as
\begin{eqnarray} 
\no^2 (t)&=&\no^2 +v_t\,[\no,\co]_+-u_t\,[\no,\so]_++v_t^2\cco+u_t^2\sso-2u_t\,v_t\,\co\so
\label{N2tCS}\\[1mm]
\langle \nno \rangle_t &=&\langle \nno \rangle_0 +
v_t\langle \, [\no,\co]_+\rangle_0-u_t\langle \,[\no,\so]_+ \,\rangle_0 
+\,v_t^2\langle \cco \rangle_0 +u_t^2\langle \sso \rangle_0
-2u_tv_t\langle \co \so \rangle_0
\label{Ntav2CS}
\end{eqnarray}
with \,$\langle\cco\rangle_0+\langle\sso\rangle_0=1$\,.
The time evolution of the variance of the position $N$ 
can then be formulated in the convenient form
\begin{equation}
\Delta^2_N(t)=\Delta^2_N(0)
+2v_t\,\Delta^2_{CN}
-2u_t\,\Delta^2_{SN}+v_t^2\,\Delta^2_{CC}+u_t^2\,\Delta^2_{SS}
-2u_tv_t\,\Delta^2_{CS}
\label{varianceNCS}
\end{equation}
where $\Delta^2_{AB}$ is the covariance of the expectation values of the operators 
$\ao$ and  $\bo$ 
at time $t=0$:
\begin{equation}
\Delta^2_{AB}=\langle\,{\textstyle \frac{1}{2}}\,[\ao,\bo]_+\,\rangle_0 
- \langle\ao\rangle_0\langle\bo\,\rangle_0\,.
\label{covarianceAB}
\end{equation}
Note that the relations for the expectation values derived above are valid for
pure states as well as for mixed states (see \cite{Goyc98a,Goyc98b,Goyc00,Goyc01} for an
application of the tight binding system using density matrices).

\section{Classicalization}

Recently, a classicalization of the tight binding model
with Hamiltonian
\begin{equation}
H=2 G(t) \cos (p \delta)+F(t)q/d \, .
\label{hamcl}
\end{equation}
has been discussed \cite{03bloch0,03bloch2D}; related observations can also be found in
\cite{Goyc00,Holt96}. 
In this classicalization, the 
operators $\no$ and $\co=(\ko+\ho)/2$ are replaced by phase space functions.
The parameter $\delta =\dd /\hbar$ depends explicitly
on $\hbar$ which implies, of course, that this 'classicalization' differs from the
usual classical limit of quantum dynamics.
It has been observed that the classical dynamics generated by (\ref{hamcl}),
\begin{equation}
\dot p =-\frac{\partial H}{\partial q}=-\frac{F}{d} \ , \quad
\dot q =\frac{\partial H}{\partial p}=-2G\delta \sin (p\delta)\,,
\label{ham-dgl}
\end{equation}
show a surprising agreement with the quantum one. Here we will analyze 
this correspondence from an algebraic point of view.

First, we can again generalize and consider a classical Hamiltonian (\ref{hamcl}) with
time dependent coefficients. Introducing the dimensionless phase space 
functions $C(p)=\cos(p \delta)$, $S(p)=\sin(p \delta)$ and $N(q)=q/\dd$, 
with Poisson brackets
\begin{equation}
\{C,N\}= +{\textstyle \frac{1}{\hbar}}\,S(p)
\ ,\quad
\{S,N\}=-{\textstyle \frac{1}{\hbar}}\,C(p)
\ ,\quad
\{C,S\}=0\,,
\label{pobraSC}
\end{equation}
the set $\{ N,C,S\}$ forms a closed Lie algebra with Lie bracket $\{\,,\,\}$ 
and therefore the dynamics 
induced by the Hamiltonian $H(p,q,t)=G(t)\,C(p)+F(t)\,N(q)$ can be
evaluated again by purely algebraic techniques. 

Moreover the classical algebra and the quantum algebra studied above are isomorphic 
in the present case which is evident from the mapping 
\begin{equation}
\no \longleftrightarrow N \ ,\quad  \co \longleftrightarrow C \ ,\quad  \so \longleftrightarrow S 
\label{mapping}
\end{equation}
where the operators $\co$ and $\so$ are defined in (\ref{CS})
(the Lie brackets map according to 
$\{\,,\,\}\longleftrightarrow
\frac{1}{\ri \hbar}\,\big[\,,\,\big]$)\,. This implies the equality of the 
dynamical evolution
of quantum operators and classical phase space functions in this case.
In particular, the evolution of the expectation values agrees.
There are, however, some differences,  e.g.~for the initial conditions. 
Whereas there is no limitation in the classical case,
the quantum covariances are limited by uncertainty relations. 

Furthermore, the equations of motion
\begin{equation}
\dot C =f S \ ,\quad \dot S =-f C \ ,\quad \dot N =-gS\,,
\end{equation}
with $f=F/\hbar$, $g=G/\hbar$, are linear which implies that the classical 
dynamics is regular, i.e.~not chaotic.

\section{The dynamical invariant}

The driven tight binding system (\ref{ham}) possesses a dynamical
invariant, very similar to the harmonic oscillator with time dependent frequency, 
where the so-called
Lewis invariant $\Io$ \cite{Lewi67,Lewi69}, a  time dependent constant of motion,
\begin{equation}
\frac{\rd \Io}{\rd t}=\frac{1}{\ri \hbar}\,\big[\,\Io,\hat H\,\big]
+\frac{\partial \Io}{\partial t}=0\,,
\label{inv-def}
\end{equation}
plays an important role. 

The dynamical algebra offers a convenient technique for constructing
such an invariant \cite{79inv,81inv}. Writing the invariant as a linear combination
of the basis of the algebra, \,$\Io =\sum_j\lambda_j \hat \Gamma_j$\,, as well as the Hamiltonian
and inserting these expressions into equation (\ref{inv-def}) using the commutator relations,
one obtains a set of linear first order differential equations for the coefficients $\lambda_j(t)$.
In the present case, we write
\begin{equation}
\Io =\gamma \no+\lambda \ko +\lambda^*\ho
\end{equation}
with $\gamma \in \mathbb{R}$ because the invariant can be chosen to be hermitian.
The procedure described above \cite{79inv,81inv} leads directly to the differential
equations
\begin{equation}
\dot \lambda = \ri\,(f_t\lambda-g_t\gamma) \, ,\quad \dot \gamma =0\,,
\end{equation}
i.e.~$\gamma$ is an arbitrary scaling constant which can be chosen as 
$\gamma=1$ and the solution for $\gamma_t$ is
\begin{equation}
\lambda_t=-\ri \re^{\ri \eta_t}\,\chi_t
\label{lambda}
\end{equation}
were $\eta_t$ and $\chi_t$ are defined in equations (\ref{USt}) and  (\ref{URt}). 
The full expression for the invariant is therefore
\begin{equation}
\Io(t)=\no(t)+\lambda_t\ko(t)+\lambda_t^*\ho(t)=\Io(0)=\no
\label{inv-t}
\end{equation}
where the time evolved operators $\ko(t)$ and $\no(t)$ have already been calculated
in equations (\ref{Kt}) and (\ref{Nt}).
The time evolution conserves the commutator relations (\ref{comm}) and therefore
$\ko(t)$ and $\ho(t)$ are still acting as ladder operators on the time dependent
eigenstates $|n,t\rangle$ of $\no(t)$ with the (time independent) eigenvalues 
$n\in \mathbb{Z}$. Following Lewis and Riesenfeld \cite{Lewi69}, one can also 
construct the  general time evolution  from the eigenstates of the
invariant.

In view of the classical version of the tight binding model in the preceding section, 
we will also express the dynamical invariant in terms of the hermitian operators
$\co$ and $\so$ given in (\ref{CS}) which yields after some algebra
\begin{equation}
\Io(t)=\no(t)+(u_t\sin \eta_t-v_t\cos \eta_t)\,\co(t) 
+(u_t\cos \eta_t+v_t\sin \eta_t)\,\so(t) \,.
\label{in-CS}
\end{equation}
In the classical version (\ref{mapping}), the invariant is the 
phase space function
\begin{equation}
I(p,q,t)=\frac{q}{d}+(u_t\sin \eta_t-v_t\cos \eta_t)\,\cos(p\delta) 
+(u_t\cos \eta_t+v_t\sin \eta_t)\,\sin(p\delta) \,.
\label{in-class}
\end{equation}
where $p=p_t$ and $q=q_t$ evolve under the Hamiltonian eqautions of motion
(\ref{ham-dgl}) and 
\begin{equation}
I(p,q,t)=I(p_0,q_0,0)=\frac{q_0}{d}
\end{equation}
is a constant of motion.

\section{Oscillating versus breathing modes and dynamic localization}
\label{sec-dynamics} 

Despite of the algebraic simplicity, the tight binding dynamics shows some non-intuitive
features, even in the case where the Hamiltonian does not explicitly depend on time,
where the famous Bloch oscillations are observed. Much more phenomena can be found for
driven system, as for instance dynamical localization effects 
\cite{Dunl86,Holt96,Grif98}.
A discussion of these phenomena is far beyond the scope of the present article. Some
quite general features, however, can be directly seen from the dynamics of the expectation
values for the position operator and its variance. For simplicity we will use the classical
description of the tight binding dynamics outlined in the preceding section.
To avoid misinterpretations, it should be recalled that the Bloch oscillation
remains, of course, a pure quantum phenomenon because the 'classical' system
is $\hbar$-dependent.

First, the general dynamical behaviour is strongly influenced by the initial distribution,
more precisely by the expectation values of $C(p)=\cos (p\delta)$,
$C^2(p)=\cos^2 (p\delta)$,\,\ldots\,
which satisfy the obvious bounds $-1\le\langle C\rangle,\langle S\rangle \le +1$ 
and $0\le\langle C^2\rangle,\langle S^2\rangle\le +1$\,. 
We will assume in the following that the initial classical phase space distribution
is symmetric in the position $q$, which implies $\langle N\rangle_0=0$ and 
$\Delta_{CN}^2= \Delta_{SN}^2=   0$, i.e.~equations 
(\ref{NtavCS}) and (\ref{varianceNCS}) read
\begin{equation} 
\langle \no \rangle_t =v_t\langle\co\rangle_0- u_t\langle\so\rangle_0
\label{NtavCS-x}
\end{equation}
\begin{equation}
\Delta^2_N(t)=\Delta^2_N(0)
+v_t^2\,\Delta^2_{CC}+u_t^2\,\Delta^2_{SS}
-2u_tv_t\,\Delta^2_{CS}
\,.
\label{varianceNCSx}
\end{equation}
Now the dynamics is most strongly influenced by the localization properties in the
momentum. Let us distinguish two extreme cases:\\[2mm]
(1) If the initial distribution is sharply localized in the vicinity of momentum $p_0$, we have
$\langle C\rangle_0   \approx \cos(p_0\delta)$, $\langle S\rangle_0\approx \sin(p_0\delta)$ and
$\Delta_{CC}^2\approx  \Delta_{SS}^2\approx \Delta_{CS}^2\approx 0$ and therefore
\begin{equation} 
\langle \no \rangle_t \approx v_t\cos(p_0\delta)- u_t\sin (p_0\delta)\,,
\end{equation}
\begin{equation}
\Delta^2_N(t)\approx \Delta^2_N(0)\,,
\end{equation}
and the distribution moves in space with constant width. This is
an oscillatory mode.\\[2mm]
(2) If the momentum distribution is broad and approximately constant over
a period of $\cos(p\delta)$, we have 
$\langle C\rangle\approx \langle S\rangle\approx 0$ and
$\Delta_{CC}^2\approx  \Delta_{SS}^2\approx 1/2$ and $\Delta_{CS}^2\approx 0$
and therefore
\begin{equation} 
\langle \no \rangle_t \approx 0\,,
\end{equation}
\begin{equation}
\Delta^2_N(t)\approx \Delta^2_N(0)+{\textstyle \frac{1}{2}}\,
\big(v_t^2+u_t^2\big)
\label{deltabr}
\end{equation}
and the distribution is frozen in space with a time dependent width. This is
a breathing mode.
(Alternatively, equation (\ref{deltabr}) can be  derived directly from 
(\ref{Utn}) for an initial distribution localized on site $n$ \cite{Dunl86}.)

In addition to these influences of the initial conditions, there are, of course, 
various effects from the time dependence. For constant fields ($g=g_0$, $f=f_0$)
$v_t=2g_0f_0^{-1}\sin (\omega_Bt)$ and $u_t=2g_0f_0^{-1}(1-\cos (\omega_Bt)\,)$
are periodic in time oscillating with the Bloch period $T_B$. Therefore
also $\langle \no\rangle_t$ oscillates with period $T_B$ in an interval of
width
$\langle \no\rangle_{\rm max}-\langle \no\rangle_{\rm min}
=4|g_0f_0^{-1}|\,\{\,\langle \co\rangle_0^2+\langle \so \rangle_0^2\,\}^{1/2}$\,.

Much more complicated is the case of an explicitly time dependent driving.
Here, we will only briefly mention the
important dynamical localization observed for resonant 
driving \,$f_t=f_0-f_1\cos (\omega t)$\, with \,$\omega_B=n\omega$\,,
$n=1,\,2,\,\ldots$\,,
(see also the discussion following equation
(\ref{fof1})\,). Here, $\chi_t$ 
grows linearly with time (\ref{Gamma}) with superimposed oscillations and 
therefore
\,$u_t\approx \gamma_nt$, $v_t \approx 0$. 
Then the dominant terms in equations (\ref{NtavCS-x}) and (\ref{varianceNCSx})
are
\begin{equation} 
\langle \no \rangle_t =- \gamma_n\,\langle\so\rangle_0\,t
\ ,\quad
\Delta^2_N(t)=\Delta^2_N(0)
+\gamma_n^2\,\Delta^2_{SS}\,t^2
\,.
\label{varianceNCSy}
\end{equation}
(valid for a space symmetric initial distribution). Therefore,
for large times, the width of the distribution increases linearly in time.
This strong dispersion can, however, be suppressed by adjusting the
field parameters to a zero of the Bessel function, i.e.
\begin{equation} 
\gamma_n =2\,g_0 \, J_n\big({\textstyle \frac{f_1}{\omega}}\big)=0\,,
\end{equation}
an effect known as dynamic localization \cite{Dunl86}
(see also \cite{Grif98,Holt96} for more details). 

\section{Single band model}

The single band model is an extension of the tight binding model by
replacing the cosine dispersion relation (\ref{Ekappa}) by a more
realistic periodic function $E(\kappa)$. Instead of the Hamiltonian
(\ref{ham}) one considers the generalization
\begin{equation}
\hat H=\hbar  \sum_{m=0}^\infty \big(\,g_m(t)\,\ko^m+g_m^*(t)\,\hmo\,\big)
+F(t)\no
=\hat H_R +F(t)\no\,.
\label{hamsb}
\end{equation}
Here, the algebra is extended to the set 
\,${\cal L}=\{\ko^m,\hmo, m\in \mathbb N, \no\}$\, with radical
\,${\cal R}=\{\ko^m,\hmo, m\in \mathbb N\}$ \,.
The subsequent analysis follows exactly the same lines as in the tight binding model. By means of 
the auxiliary relations \,$[\ko^m\,,\,\no\,]=2^{m-1}\ko^m$\, and
\,$[\hmo\,,\,\no\,]=-2^{m-1}\hmo$\, one obtains
\begin{equation}
\hat U_S^{-1}\hat H_R\,\hat U_S=\sum_{m=0}^\infty \big(\,g_m(t)\,\re^{-\ri 2^{m-1}\eta_t\,\no}
\,\ko^m +g_m^*(t)\,\re^{+\ri 2^{m-1}\eta_t\,\no }\,\hmo \,\big)
\end{equation}
and the time evolution of the radical part of the algebra is given by
\begin{equation}
\hat U_R(t)=\exp \Big({\textstyle -\ri \sum_{m=0}^\infty
\big(\,\chi_m(t)\,\ko^m+\chi_m^*(t)\,\hmo\,\big)}\Big)
\label{URsb}
\end{equation}
with
\begin{equation}
\chi_m(t)=\int_0^t\rd \tau\, g_m(\tau)\,\re^{-\ri 2^{m-1}\eta_\tau}\,.
\end{equation}
The full time evolution operator is again
\,$\hat U(t)=\hat U_S(t)\,\hat U_R(t)$\,, where $\hat U_S(t)$ is still 
given by (\ref{USt}). 

For most applications, however, the coefficients $g_m$ will be independent
of time. In such a case, the dispersion relation
is given by the Fourier series
\begin{equation}
E(\kk)=\hbar \sum_{m=0}^\infty \big(\,g_m\,\re^{\ri m \kk}+g_m^*\,\re^{-\ri m \kk}\,\big)\,.
\label{dispsb}
\end{equation}

Matrix elements of \,$\hat U(t)$\, in the Bloch basis are similar
to  (\ref{Utkappa}):
\begin{equation}
\langle \kk|\hat U(t)|\kk'\rangle=
\delta_{2\pi}(\kk -\kk'-\eta_t)\,\re^{-\ri \,(\,f(\kk,t)+f^*(\kk,t)\,)}\,,
\label{Utkappa-sb}
\end{equation}
with \,$f(\kk,t)= \sum_m\chi_m(t)\re^{\ri m\kk}$\,.
The matrix elements in the $|n\rangle$ basis are, however, more complicated 
than the tight binding expression (\ref{Utn}).

The time evolution of the ladder operator $\ko(t)$ is still given by 
(\ref{Kt}) and, using
\begin{equation}
\re^{\,z\ad \ko^m}\,\no=\no+z\,2^{m-1}\,\ko^m\,,
\end{equation}
the evolution of the position operator is
\begin{equation}
\no(t)=\no+\ri \sum_{m=0}^\infty 2^{m-1}\big(\, \chi_m(t)\,\ko^m-\chi_m^*(t)\,\hmo\,\big)
\end{equation}
with expectation value
\begin{equation}
\langle \no\rangle_t=\langle \no\rangle_0+\ri \sum_{m=0}^\infty 2^{m-1}
\big(\, \chi_m(t)\,\langle \ko^m\rangle-\chi_m^*(t)\,\langle \hmo\rangle\,\big)\,.
\end{equation}
Finally it should be noted that also in this case a
classicalization is possible with classical Hamiltonian
\begin{equation}
H(p,q,t)=E(p\delta)+Fq
\end{equation}
as discussed above for the tight binding model.

\section{Concluding remarks}

The driven tight binding system is in many aspects very similar to
the driven harmonic oscillator. It can be treated algebraically
by means of ladder operators, there exists a dynamical invariant,
and one observes a close correspondence between quantum
and classical time evolution. Some
of these features of the tight binding dynamics have been
discussed in the present paper. There are, of course, still
a number of interesting questions to be answered as for
instance the role of the coherent states \cite{Kowa96,Gonz98,Kowa02}
and 
the relation between the invariant and the quasienergies for
a periodically driven system \cite{94inv,98dyninv}. Moreover, an extension
of the algebraic technique to treat Bloch-Zener oscillations in 
doubly periodic structures \cite{03bloch2D} or the recently
investigated 
two dimensional case \cite{02tb2d,03bloch2D,Dmit01-02a,03bloch1}.
Work in these directions is in progress.


\end{document}